\def\frac#1#2{{{{#1}}\over{{#2}}}}
\documentclass[12pt]{article}
\usepackage{amsmath}
\usepackage{wasysym}    
\usepackage{textcomp}
\usepackage{pict2e}
\usepackage{cite}

\usepackage{xcolor, cancel, ulem}
\usepackage{graphicx}
\usepackage{hyperref}
\usepackage{pdfpages}
\usepackage{cite}

\newsavebox{\ns}
\newsavebox{\dbrane}
\newsavebox{\dbshort}

\usepackage{epsfig}
\usepackage{amssymb}

\def\appendix{{\newpage\section*{Appendix}}\let\appendix\section%
        {\setcounter{section}{0}
        \gdef\thesection{\Alph{section}}}\section}

\newcommand\ba{\begin{eqnarray}}
\newcommand\ea{\end{eqnarray}}

\definecolor{DarkGreen}{rgb}{0,.64,0}
\definecolor{gunmetal}{rgb}{0.171875, 0.207031, 0.222656}
\definecolor{chartreuse}{rgb}{.49,.98,0}
\definecolor{amethyst}{rgb}{0.59375,0.398438,0.792969}
\definecolor{brownrust}{rgb}{0.6875, 0.316406, 0.242188}
\definecolor{Violet}{rgb}{0.5,0,1}
\definecolor{BurntOrange}{rgb}{0.792969,0.332031,0}
\definecolor{FreshEggplant}{rgb}{0.59375, 0., 0.414063}
\definecolor{salmon}{rgb}{0.996094,0.507813,0.410156}
 \definecolor{FrenchRose}{rgb}{0.96875, 0.292969, 0.5625}
\definecolor{Cabaret}{rgb}{0.808594, 0.242188, 0.46875}
\definecolor{Shamrock}{rgb}{0.242188, 0.808594, 0.582031}
\definecolor{RobinsEggBlue}{rgb}{0., 0.792969, 0.792969}
\definecolor{GuardsmanRed}{rgb}{0.792969, 0., 0.}
\definecolor{Sapphire}{rgb}{0.183594, 0.328125, 0.621094}
\definecolor{Sorbus}{rgb}{0.996094, 0.429688, 0.0273438}
\definecolor{Red}{rgb}{1,0,0}
\definecolor{Blue}{rgb}{0,0,1}
\definecolor{Black}{rgb}{0,0,0}
\definecolor{Green}{rgb}{0,1,0}
\definecolor{thistle3}{rgb}{0.800781, 0.707031, 0.800781}
\definecolor{thistle4}{rgb}{0.542969, 0.480469, 0.542969}
\definecolor{DarkTurquoise}{RGB}{0,206,209}
\definecolor{turquoise4}{RGB}{0,134,139}

\definecolor{Purple}{rgb}{0.808594, 0.242188, 0.46875}
\newcommand{\purple}[1]{{\color{Purple} {#1}}}

\newcommand{\TENSOREDIT}[1]{{}}

\newcommand{\nn}{\nonumber}

\def\Dslash{\,\,{\raise.15ex\hbox{/}\mkern-12mu D}}
\def\Dbarslash{\,\,{\raise.15ex\hbox{/}\mkern-12mu {\bar D}}}
\def\delslash{\,\,{\raise.15ex\hbox{/}\mkern-9mu \partial}}
\def\delbarslash{\,\,{\raise.15ex\hbox{/}\mkern-9mu {\bar\partial}}}
\def\pslash{\,\,{\raise.15ex\hbox{/}\mkern-9mu p}}
\def\calDslash{\,\,{\raise.15ex\hbox{/}\mkern-12mu {\cal D}}}

\newcommand{\hh}{{1\over 2}}

\renewcommand{\ll}{_}
\newcommand{\uu}{^}
\newcommand{\pp}{\partial}

\renewcommand{\exp}[1]{{\rm exp}\{#1\}}

\renewcommand{\d}{\delta}
\newcommand{\m}{\mu}

\renewcommand{\m}{\mu}
\newcommand{\n}{\nu}
\newcommand{\s}{\sigma}

\renewcommand{\o}{\omega}
\newcommand{\e}{\epsilon}

\newcommand{\sqd}{^2}

\def\wb{{\bar{w}}}

\newcommand{\pb}{{\bar{\partial}}}

\renewcommand{\hh}{{1\over 2}}
\renewcommand{\gg}{\nabla}

\newcommand{\eee}[1]{\ba{#1}\ea}

\renewcommand{\b}{\beta}

\newcommand{\llsk}{\hskip .5in}

\newcommand{\D}{\Delta}

\newcommand{\apr}{{\alpha^\prime} {}}

\newcommand{\IZ}{\relax\ifmmode\mathchoice
{\hbox{\cmss Z\kern-.4em Z}}{\hbox{\cmss Z\kern-.4em Z}}
{\lower.9pt\hbox{\cmsss Z\kern-.4em Z}} {\lower1.2pt\hbox{\cmsss
Z\kern-.4em Z}}\else{\cmss Z\kern-.4em Z}\fi} \font\cmss=cmss10
\font\cmsss=cmss10 at 7pt
\newcommand{\inbar}{\,\vrule height1.5ex width.4pt depth0pt}
\newcommand{\IC}{{\relax\hbox{$\inbar\kern-.3em{\rm C}$}}}
\newcommand{\IQ}{{\relax\hbox{$\inbar\kern-.3em{\rm Q}$}}}
\newcommand{\IP}{\relax{\rm I\kern-.18em P}}

\newcommand{\ed}{\dot{e}}

\newcommand{\ca}{c_{(\alpha)}}

\newcommand{\cc}{{\cal C}}

\renewcommand{\cc}{{c_1}}

\newcommand{\cm}{{\cal M}}

\newcommand{\ups}[1]{^{[{#1}]}}

\renewcommand{\o}{\omega}

\renewcommand{\cc}{c}

\newcommand{\pr}{^\prime}

\newcommand{\IR}{\relax{\rm I\kern-.18em R}}
\def\blfootnote{\xdef\@thefnmark{}\@footnotetext}

\renewcommand{\cc}[1]{\cite{#1}}

\renewcommand{\ca}{{\cal A}}
\newcommand{\bm}{\begin{matrix}}
\renewcommand{\em}{\end{matrix}}

\newcommand{\ee}[1]{\ba {#1} \ea}

\newcommand{\upp}[1]{^{({#1})}{}}

\newcommand{\co}{{\cal O}}

\newcommand{\rr}[1]{(\ref{{#1}})}
\newcommand{\bbb}{\ba\begin{array}{c}}
\renewcommand{\eee}{\nonumber\end{array}\ea}
\newcommand{\een}[1]{\label{#1}\end{array}\ea}

\renewcommand{\cm}{{\cal M}}

\def\ca{{\cal A}}

\def\uth{^{\rm{\underline{th}}}}

\def\bi{\begin{itemize}}
\def\ei{\end{itemize}}

\def\ed{\end{document}}

\def\cc{{\cal C}}

\renewcommand{\rr}[1]{(\ref{#1})}

\def\cc{\,}

\def\cc{\,}

\newcommand{\lp}[1]{_{({#1})}}

\newcommand{\lrm}[1]{_{{\rm {#1}}}}
\newcommand{\urm}[1]{^{{\rm {#1}}}}
\newcommand{\uprm}[1]{^{({\rm {#1}})}}

\newcommand{\ls}[1]{_{[{#1}]}}

\usepackage{fancyhdr}
\renewcommand{\eqref}[1]{\rr{#1}}
\usepackage{dsfont}

\newcommand{\redd}[1]{{\color{Red} {#1}}}

\definecolor{Purple}{rgb}{0.808594, 0.242188, 0.46875}

\newcommand{\p}{\partial}
\newcommand{\bp}{\bar\partial}

\newcommand{\al}{\alpha'}
\newcommand{\eii}{E_{11}}

\def\xxn{\\ \\ {}}

\renewcommand{\bm}{\begin{matrix}}
\renewcommand{\em}{\end{matrix}}

\newcommand{\aaa}[1]{}

\renewcommand{\caption}[1]{\begin{center}{{#1}}\end{center}}

\def\be{\begin{eqnarray}}
\def\ee{\end{eqnarray}}
\def\nn{\nonumber}

\newcommand{\WhiteOutWithNotification}[1]{ {  \color{Red}  (  A SECTION HAS BEEN WHITED OUT HERE. \cc   )  } }
\newcommand{\WhiteOut}[1]{}
\newcommand{\RoundIEdit}[1]{}
\newcommand{\RoundHEdit}[1]{}
\newcommand{\RoundGEdit}[1]{}
\newcommand{\RoundFEdit}[1]{}
\newcommand{\RoundEEdit}[1]{}
\newcommand{\EMVertEdit}[1]{}
\newcommand{\HugeEditB}[1]{}
\newcommand{\HugeEditA}[1]{}
\newcommand{\RoundDEdit}[1]{}
\newcommand{\RoundCEdit}[1]{}
\newcommand{\RoundBEdit}[1]{}
\newcommand{\RoundAEdit}[1]{}

\newcommand{\blue}[1]{{\color{Blue}{#1}}}

\def\redlowdash{{\color{Red}{\rule[-0.5ex]{2pt}{0.4pt}}}}
\def\redmiddash{{\color{Red}{\rule[+0.5ex]{2pt}{0.4pt}}}}

\def\cute{{\lower3.5pt\hbox{\sixly
  \kern-.21pt \char58 \kern-.21pt }}}

\def\midcute{{\lower-1.0pt\hbox{\sixly
  \kern-.21pt \char58 \kern-.21pt }}}

  \def\lowcute{{\lower3.5pt\hbox{\sixly
  \kern-.21pt \char58 \kern-.21pt }}}
  
  \def\redmidcute{{\color{Red} \midcute}}
  
  \def\redlowcute{{\color{Red} \lowcute}}
   
    \def\bluelowcute{{\color{Blue} \lowcute}}

   \def\swave{\bgroup \markoverwith \midcute \ULon} 
  
  \def\redswave{\bgroup \markoverwith \redmidcute \ULon} 
  
  \def\reduline{\bgroup \markoverwith \redlowdash \ULon}
  
   \def\blueuline{\bgroup \markoverwith \bluelowdash \ULon}
  
   \def\reduwave{\bgroup \markoverwith \redlowcute \ULon}
   
   \def\blueuwave{\bgroup \markoverwith \bluelowcute \ULon}
  
  \def\redsout{\bgroup \markoverwith \redmiddash \ULon}
  
   \def\bluesout{\bgroup \markoverwith \bluemiddash \ULon}
   
   \def\Irrel{\bgroup \markoverwith {{\color{Red} {\bf X}}} \ULon}

\newcommand{\eqirrel}[1]{\rdots}
  
\let\oldcancel\cancel
\renewcommand\cancel[1][black]{%
  \def\CancelColor{\color{#1}}%
  \oldcancel}

   \let\oldbcancel\bcancel
\renewcommand\bcancel[1][black]{%
  \def\CancelColor{\color{#1}}%
  \oldbcancel}

\def\bca{\begin{cases}}
\def\eca{\end{cases}}

\def\rdots{\redd{\circ\circ\circ}}

\def\hinv{H}

\renewcommand{\upp}[1]{^{({#1})}}

\def\lsim{\mathrel{\lower0.3em\hbox{$\stackrel{\textstyle <}{\sim}$}}}
\def\gsim{\mathrel{\lower0.3em\hbox{$\stackrel{\textstyle >}{\sim}$}}}

\def\negspace{\kern -0.4em}

\def\dvec{\raise 0.3 em \hbox{$^\leftrightarrow$} \kern -0.77 em}

\def\omegahat{\hat%
	{\setbox0=\hbox{$\omega$}%
		\kern-.025em\copy0\kern-\wd0
		\kern.05em\copy0\kern-\wd0
		\kern-.025em\raise.0433em\box0}}

\def\cM{{\cal M}}

\def\pol#1{}
\def\cald{{\cal D}}

\def\inv{X}
\def\inv{\blue{\tt{Inv}}}
\def\inv{{\blue{{\cal I}}}}

\def\phc{{\purple{\varphi}}}
\newcommand{\putat}[3]{\begin{picture}(0,0)(0,0)\put(#1,#2){#3}\end{picture}}

\def\SilentOthersComment#1{{}}

\newcommand{\defout}[1]{}

\begin{document}

\begin{titlepage}
\begin{flushright}
IPMU-14-0125\\
\end{flushright}
\vspace{8 mm}
\begin{center}
  {\Large \bf Effective String Theory \\ \vspace{4 mm} Simplified }
\end{center}
\vspace{2 mm}
\begin{center}
{Simeon Hellerman$^{1,a}$, Shunsuke Maeda$^{1,2,b}$, Jonathan Maltz$^{1,c}$, Ian Swanson$^d$}\\
\vspace{6mm}
{\it $^1$Kavli Institute for the Physics and Mathematics of the Universe\\
The University of Tokyo \\
 Kashiwa, Chiba  277-8582, Japan\\}
 \vspace{6mm}
{\it $^2$Department of Physics, Faculty of Science,\\
University of Tokyo, Bunkyo-ku, Tokyo 133-0022, Japan\\}
\vspace{5mm}
{\tt  $^a$simeon.hellerman.1@gmail.com, $^b$shunsuke.maeda@ipmu.jp,
 $^c$jdmaltz@alumni.stanford.edu, $^d$ianswanson.physics@gmail.com}
 \vspace{6mm}

\end{center}
\vspace{-10 mm}
\begin{center}
{\large Abstract}
\end{center}
\noindent

In this note we simplify the formulation of the Poincar\'e-invariant effective string theory in $D$ dimensions by
adding an intrinsic metric and embedding its dynamics into the Polyakov formalism.
We use this formalism to classify operators order-by-order in the inverse physical length of the string,
in a fully gauge-invariant framework.  We then use this classification to analyze 
the universality and nonuniversality of observables, up to and including 
the second sub-leading order in the long string expansion.

\vspace{1cm}
\begin{flushleft}
\today
\end{flushleft}
\end{titlepage}

\tableofcontents
\newpage

\section{Introduction} 
String theory was originally developed as a tool to study the dynamics of chromoelectric flux tubes in
strongly coupled gauge theory\RoundAEdit{, but it quickly got sort of sidetracked into the study of gravity}.  Some time
later, a consistent quantization was developed \cite{Polchinski:1991ax} for Poincar\'e-invariant string theories in $D \neq 26$, in the 
limit where the physical scale of the string is much larger than the square root of the inverse
string tension, or $\sqrt{\apr}$.  Decades after the original discovery of string theory,
the study of gravity led back to the application of string theory to strongly coupled 
gauge theory, via gauge-gravity duality and the holographic principle \cite{Aharony:1999ti}.  
The relationship between
holographic string duals of confining gauge theories and effective non-gravitational string theory was
to some extent worked out in \cite{Aharony:2013ipa}, drawing on earlier work 
on the derivation of effective string theories from fundamental strings propagating on warped geometries \cite{natsuume} 
and perturbed Liouville theories \cite{Polchinski:1991ax}.

The effective theory of relativistic strings has developed in spite of the lack of a truly simple formalism.  
Static gauge \cite{Luscher:1980fr, Luscher:1980ac, Luscher:2004ib} lacks manifest 
covariance, and calculations in static gauge at the quantum
level are vulnerable to subtle errors due to Lorentz-breaking effects entering through the procedures used
to regularize and renormalize the theory. The Polchinski-Strominger (PS) formalism, while manifestly relativistically covariant
and relatively easy to quantize, is based on the ad hoc addition of a singular interaction to the
gauge-fixed Lagrangian.  

Therefore, it is desirable to develop a more systematic treatment of the quantization of string dynamics in the
effective framework prior to gauge fixing, so that the translation
between different gauges and renormalization schemes 
can be carried out with ease and clarity.  In this paper we develop such an
approach, by embedding the effective string into the Polyakov
formalism.  (We also direct the reader's attention to the 
earlier work \cite{HariDass:2007gn}, where an embedding of the effective string into the Polyakov path integral was
also pursued.  We thank N. D. Hari Dass for making us aware of this work.)

\section{Effective string theory in the Polyakov formalism }
\subsection{Generalities}
The Polyakov string is defined by the path integral
\bbb
Z = \int \cc \cald\m\ls g\urm{Polyakov}\cc \exp{- S\ll{{\rm Polyakov}}}\ ,
\xxn
\cald\m\ls g\urm{Polyakov} \equiv {{\cald\ls g X \cc \cald\ls g g }\over{\cald\ls g \Omega}} \cc \ ,
\xxn
S\lrm{Polyakov} = \int \cc d\sqd \s \cc {\sqrt{|g\ll{\bullet\bullet}|}}{\cal L}\lrm{Polyakov}\ ,
\xxn
 {\cal L}\lrm{Polyakov} = {1\over{4\pi\apr}} \cc g\uu{ab}\cc \pp\ll a X\uu\m \cc \pp\ll b X\ll\m\ .
\eee
The subscript $[g]$ on the various factors of the path integral measure mean that the various pieces of the
measure are regularized and renormalized with a local prescription using the fiducial metric $g\ll{\bullet\bullet}$.

The action $S\lrm{Polyakov}$ is Weyl-invariant, but the measure $\cald\cm\ls g\urm{Polyakov}$ is not,
transforming under Weyl transformations $g\pr\ll{\bullet\bullet} = \exp{2\o} \cc g\ll{\bullet\bullet}$ as
\bbb
\cald\cM\ls {g\pr}\urm {Polyakov} = \exp{ {{D-26}\over{24\pi}}\int \cc d\sqd\s \cc \sqrt{|g|} \cc\big ( g\uu{ab} \cc  \pp\ll a\o \cc
\pp\ll b \o  + \o\cc {\cal R}\lp 2 \big )} \cdot \cald \cM\ls {g}\urm{Polyakov}\ ,
\een{AnomalousMeasureTransformation}
in a general number of dimensions $D$.  
The expression above is based on a Euclidean-signature
worldsheet.  For a Lorentzian-signature worldsheet metric $g$ (with $g\ll{00} < 0$ signature convention), we have
\bbb
\cald\cM\ls {g\pr}\urm {Polyakov} = \exp{i {{D-26}\over{24\pi}}\int \cc d\s\uu 0\cc d\s\uu 1 \cc \sqrt{|g|} \cc \left ( g\uu{ab} \cc  \pp\ll a\o \cc
\pp\ll b \o  +\o\cc {\cal R}\lp 2 \right )} \cdot \cald \cM\ls {g}\urm{Polyakov}\ .
\eee

In the ``linear dilaton" or ``quantum Liouville theory" approach, one cancels this anomaly by assigning a nontrivial Weyl transformation to one
of the scalars $X\uu{D-1} \equiv {1\over{|V|}} \cc V\ll\m X\uu\m$:
\bbb
X\uu{D-1} \to X\uu{D-1} - \sqrt{{{26-D}\over{6\apr}}} \cc\o\ .
\eee
Then $X$ can be related to the Liouville field $\phi$, which transforms under Weyl transformations with a unit shift, by
\bbb
X\uu{D-1} = - \sqrt{{26-D}\over{6\apr}}\cc \phi\ , \llsk\llsk \phi =   - \sqrt{{6\apr}\over{26-D}}\cc X\uu{D-1} \ .
\eee
The transformation of $\phi$ is thus
\bbb
\phi\to\phi + \o\ ,
\eee
and the action for $\phi$ (in Euclidean signature) is
\bbb
S\ll\phi =  {{26-D}\over{24\pi}}\int \cc d\sqd\s \cc \sqrt{|g|} \cc \left ( \cc  g\uu{ab} \cc  \pp\ll a\phi \cc
\pp\ll b \phi  -\phi\cc {\cal R}\lp 2 \right )\ ,
\een{LiouvilleAction}
or
\bbb
S\ll\phi =  {{D-26}\over{24\pi}}\int \cc d\s\uu 0 \cc d\s\uu 1 \cc \sqrt{|g|}   \cc \left (  \cc  g\uu{ab}\pp\ll a\phi \cc
\pp\ll b \phi  - \phi\cc {\cal R}\lp 2 \right )
\eee
in Lorentzian signature ($g\ll{00} < 0$).  
The anomalous transformation of the measure is then cancelled exactly
by the classical transformation of the action for $\phi$.  The latter depends only on the transformation
of the $\phi$ field itself.  Therefore, we can achieve the exact same classical transformation of the action
by substituting any composite operator $\phc$ for $\phi$ that transforms as
\bbb
\phc \to \phc + \o
\een{PhiCompTrans}
under a Weyl transformation, and transforms as a scalar under diffeomorphisms.
One can easily construct such operators from $X\uu\m$ and $g\ll{ab}$, while leaving all $X\uu\m$ to transform
trivially under Weyl transformations, thus preserving $D$-dimensional Poincar\'e invariance.  The simplest
scalar operator that transforms as \rr{PhiCompTrans}, is
\bbb
\phc \equiv - \hh \cc {\rm ln}(g\uu{ab} \pp\ll a  X\uu\m\cc\pp\ll b X\ll\m ) \ .
\een{phiCompDef}
So, in terms of $\phc$, our action becomes
\bbb
S = S\ll{{\rm Polyakov}} + S\ll{{  {{\rm composite}\atop{\rm Liouville}}}}\ ,
\een{EffectiveStringFullBareAction}
where
\bbb
S\ll{{  {{\rm composite}\atop{\rm Liouville}}}} \equiv  S\ll\phc =  {\b\over{2\pi}}\int \cc d\sqd\s \cc \sqrt{|g|} \cc \left ( \cc  g\uu{ab} \cc  \pp\ll a\phc \cc
\pp\ll b \phc  -\phc\cc {\cal R}\lp 2 \right ) \ ,
\een{CompositeLiouvilleAction}
where we have defined the coefficient
\bbb
\b\equiv {{26- D}\over{12}}\ .
\een{BetaValue}
We have written the coefficient of the anomaly
action in terms of $\b$ to connect later on with the conventions
and notation \cite{Polchinski:1991ax}
of the old covariant formalism.\footnote{Note that there is a sign error in the equation expressing
the value of $\b$ in  \cite{Polchinski:1991ax}.  The sign of $\b$ in our \rr{BetaValue} is
the correct one, and is in agreement with all calculations and equations of \cite{Polchinski:1991ax}, other than their equation (15).
The incorrect sign in that expression appears to be a simple typographical error.}

As noted, with this definition of the action, the anomalous 
transformation \rr{AnomalousMeasureTransformation} of the measure 
$ \cald \cM\ls {g}\urm{Polyakov}$ is precisely cancelled by the classical transformation of 
the action \rr{EffectiveStringFullBareAction}, and the path
integral is invariant.  Of course there are many (indeed, an infinite number of) additional 
diff $\times$ Weyl-invariant terms one could add to the action, and, correspondingly, many (an infinite number of)
ambiguities in the definition of the theory arising from dependence on the regularization 
and renormalization scheme.  These two sets of ambiguities are of course
the same: Any two local regularization and renormalization procedures must differ in effect by local 
terms in the bare action.  Any two renormalization
schemes that properly preserve all the symmetries, including diff $\times$ Weyl invariance, must differ 
by local terms that preserve the same
symmetries.  However, there is no local term preserving all symmetries that scales as $|X|\uu 0$ in the long 
string expansion, other than the Euler density, which is
topological.  Thus, any two consistent renormalization schemes must yield equivalent amplitudes for 
all processes at first subleading order in the long-string expansion,
modulo a possible renormalization of the effective string coupling for processes involving scattering or decays.  
In particular, there can be no disagreement among 
observable processes for different ``quantizations" of the effective string at first subleading order, so 
long as the symmetries are properly preserved.\footnote{For a beautiful analysis
of matching between various gauges and renormalization schemes, the interested reader 
is urged to consult \cite{Aharony:2013ipa, Dubovsky:2012sh}.}

\subsection{Stress tensor in the effective string in Polyakov formalism}
The major difference between the linear dilaton case and the composite-Liouville case is that 
the composite Liouville field $\phc$ itself involves
the metric in its definition.  That is, suppose we vary the metric by an infinitesimal amount
 \bbb
g\uu{ab} \to g\uu{ab} + h\uu{ab}, \llsk g\ll{ab} \to g\ll{ab} - h\ll{ab}\ , \llsk h\ll{ab} \equiv g\ll{ac} g\ll{bd} h\uu{cd}\ ,
\eee
whereby ${\d\over{\d h\uu{ab}}}\cc \phc \neq 0$.  Note that this is in contrast to the case 
of a conventional Liouville field $\phi$, which is an independent
degree of freedom defined without respect to the metric.  In particular, with the definition \rr{phiCompDef}, 
we have
\bbb 
\frac {\delta \phc}{\delta h^{ab}}=-\frac{1}{2}\frac{\partial_a X \cdot \partial_b X}{(\nabla X)^2}\ .
\een{phiCompIntrinsicVar}
We can thus write expressions for the variation of the 
kinetic term and the Ricci coupling for the composite Liouville field $\phc$.  
To this end, we apply standard formulae from
differential geometry, remembering to supplement terms from the explicit metric
variation of the action at fixed $\phc$ with terms coming from the metric 
variation of $\phc$ itself, shown in eqn.~\rr{phiCompIntrinsicVar}.

Defining
\bbb
S\uprm{kin} \equiv {{26-D}\over{24\pi}} \cc \int \cc  {\sqrt{|g|}} \cc d\sqd \s \cc  g\uu{ab} \pp\ll a \phc \pp\ll b \phc\ ,
\eee
and
\bbb
S\uprm{Ricci} \equiv  - {{26-D}\over{24\pi}}\int \cc d\sqd\s \cc \sqrt{|g|} \cc \phc \cc {\cal R}\lp 2\ ,
\eee
we find the contributions to the stress tensor at order $\b\uu 1 \cc |X|\uu 0$ by taking the variation of the 
order-$\b\uu 1 \cc |X|\uu 0$ piece of the worldsheet action, which is just the anomaly-cancelling term $S\uprm{kin} + S\uprm{Ricci}$.
Defining\footnote{In accordance with the sign and normalization
conventions 
of \cite{Polchinski:1991ax} for the stress tensor.} $ T\ll{ab}^{[\beta^1|X|^0]} \equiv -  4\pi{{\d }\over{\d h\uu{ab}}}S\ll{{  {{\rm composite}\atop{\rm Liouville}}}}$ 
and using \rr{phiCompIntrinsicVar}, we find 
\bbb
 T\ll{ab\RoundCEdit{~(\phc)}}\ups{\b\uu 1 |X|\uu 0}
  = T\uprm{kin}\ll{ab\RoundCEdit{~(\phc)}} + T\uprm{Ricci}\ll{ab\RoundCEdit{~(\phc)}}\ ,
\eee
where
\be
 T\uprm{kin}\ll{ab\RoundCEdit{~(\phc)}} 
 &=& {{26-D}\over 6} \cc \bigg [ -(\gg\ll a \phc \gg\ll b\phc - \hh \cc g\ll{ab} \cc (\gg \phc)\sqd) 
-  (\gg\sqd \phc ) \big ( +  {{ \gg\ll a X \cdot \gg\ll b X}\over{ (\gg X)\sqd}}   \big )\bigg ]\ ,
 \nn\\
 T\uprm{Ricci}\ll{ab\RoundCEdit{~(\phc)}}
 &=& 
 {{D-26}\over 6}\cc \bigg [ \cc \gg\ll a \gg\ll b \phc - g\ll{ab} \gg\uu c \gg\ll c \phc
+ \hh {{ \pp\ll a X \cdot \pp\ll b X}\over{ (\gg X)\sqd}} \cc {\cal R}\lp 2  \cc \bigg ] \ .
\label{TotalInteractingStressPieces}
\ee
The trace of the two pieces above appear (classically) as
\bbb
T\uprm{kin}\ll{ \RoundFEdit{   (  \phc  )     }  }{}\uu a{}\ll a= g\uu{ab} \cc  T\uprm{kin}\ll{ab\RoundFEdit{~(\phc)}} = - {{26-D}\over 6} \cc
\gg\sqd\phc\ ,
\xxn
T\uprm{Ricci}\ll{ \RoundFEdit{   (  \phc  )  }    }{}\uu a{}\ll a= g\uu{ab} \cc  T\uprm{Ricci}\ll{ab\RoundFEdit{~(\phc)}} =
+ {{26-D}\over 6} \cc
\gg\sqd\phc - {{26 - D}\over{12}} \cc {{\cal R}}\lp 2\ .
\eee
So the trace of the total order-$\b\uu 1 |X|\uu 0$ stress tensor (classically) is
\bbb
T\ups{\b\uu 1 |X|\uu 0}\ll{\RoundCEdit{(\phc)}}{}\uu a{}\ll a
 =
 - {{26 - D}\over{12}} \cc {{\cal R}}\lp 2\ ,\llsk{({\rm modulo~quantum~corrections})}\ .
\eee
The quantum correction, given by equation (3.4.15) of \cite{Polchinski:1998rq}, is
\bbb
T\uprm{quantum}\ll a{}\uu a = - {c\over{12}}\cc {{\cal R}}\lp 2\ ,
\eee
where, for us, $c = {{D-26}}$.  We then have
\bbb
T\uprm{quantum}\ll a{}\uu a = {{26-D}\over{12}} \cc {{\cal R}}\lp 2\ .
\eee
Thus, the total (classical plus quantum) trace of the stress tensor is
\bbb
T\uu a{}\ll a\big |\upp{{\rm classical}, ~O(\b\uu 1 |X|\uu 0)}\ll{\RoundCEdit{(\phc)}} + T\uu a{}\ll a\big |\uprm{quantum}= 0\ .
\eee

For a flat metric in Euclidean signature, $g\ll{ab} = \d\ll{ab}$,
 meaning $\d\ll{ww} = \d\ll{\wb\wb} = 0$ and $\d\ll{w\wb} = \d\ll{\wb
 w} = \hh$ in the standard complex coordinates
\be
w &\equiv& \s\uu 2 + i \s\uu 1\ ,
\nn\\
\wb &\equiv& \s\uu 2 - i \s\uu 1\ .
\ee
Then ${\cal R}\lp 2 = 0$ and we have
\bbb
T\uprm{kin}\ll{ww~\RoundFEdit{(\phc)}}  = {{26-D}\over 6} \cc \bigg [ -\pp\ll w \phc \pp\ll w\phc 
-  (\gg\sqd \phc ) \big (   {{ \pp\ll w X \cdot \pp\ll w X}\over{ (\gg X)\sqd}}   \big )\bigg ]\ ,
\xxn
 T\uprm{Ricci}\ll{ww~\RoundFEdit{(\phc)}} = - {{26-D}\over 6}\cc  \pp\sqd\ll w \phc \ .
\een{FlatStressTensorInteractionMetricVariationMethodMysteriouslyIncorrectNonConserved}
Then, using
\bbb
\gg\sqd = 4 \cc \pp\ll w \pp\ll \wb
\eee
and
\bbb
\gg X \cdot \gg X = 4\cc  \pp\ll w X \cdot \pp\ll\wb X\ ,
\eee
we have
\bbb
T\uprm{kin}\ll{ww\RoundFEdit{~(\phc)}}  = {{26-D}\over 6} \cc \bigg [ -\pp\ll w \phc \pp\ll w\phc 
+  (\pp\ll w \pp\ll\wb \phc ) \big (   {{ \pp\ll w X \cdot \pp\ll w X}\over{ \pp\ll w X\cdot \pp\ll\wb X}}   \big )\bigg ]\ ,
\xxn
 T\uprm{Ricci}\ll{ww\RoundFEdit{~(\phc)}} = - {{26-D}\over 6}\cc  \pp\sqd\ll w \phc \ .
\eee
For brevity, we write $\pp\ll w \equiv \pp$ and $\pp\ll \wb\equiv \pb$, so that
\bbb
T\uprm{kin}\ll{ww\RoundFEdit{~(\phc)}}  = {{26-D}\over 6} \cc \bigg [ -(\pp \phc)\sqd
+  (\pp\pb \phc ) \big (   {{ (\pp X)\sqd }\over{ \pp X \cdot \pb X}}   \big )\bigg ]\ ,
\xxn
 T\uprm{Ricci}\ll{ww\RoundFEdit{~(\phc)}} = - {{26-D}\over 6}\cc  \pp\sqd \phc \ .
\een{StressTensorPiecesFlat}
For $g\ll{ab} = \d\ll{ab}$ we have $\phc = - \hh \cc {\rm ln}( 4\cc  \pp X \cdot \pb X) = - \hh \cc {\rm ln}(\pp X \cdot \pb X) + {\rm const.}$.  So,
discarding terms proportional to the leading order equations of motion 
(that is, setting $\pp\pb X = O(\b X\uu{-1})$ by the EOM), we have
\bbb
\pp\phc = - \hh \cc {{\pp\sqd X \cdot \pb X}\over{\pp X\cdot \pb X}}\ ,
\xxn
\pp\sqd\phc = - \hh \cc{{\pp\uu 3 X \cdot \pb X}\over{\pp X \cdot \pb X}} + \hh \cc {{(\pp\sqd X \cdot \pb X)\sqd}\over{(\pp X \cdot \pb X)\sqd}}\ ,
\eee
and
\bbb
\pp\pb \phc = - \hh \cc {{\pp\sqd X \cdot \pb\sqd X}\over{\pp X\cdot \pb X}} + \hh \cc  {{(\pp\sqd X \cdot \pb X)(\pp X \cdot \pb\sqd X)}\over{(\pp X\cdot \pb X)\sqd}}\ .
\eee
Using the notation $\inv\ll{pq} \equiv \pp\uu p X\cdot \pb\uu q X$ in Euclidean signature 
(or $\inv\ll{pq} \equiv \pp\ll +\uu p X\cdot \pp\ll -\uu q X$ for
Lorentzian signature), we have
\bbb
\pp\phc = - \hh \cc {{\inv\ll{21}}\over{\inv\ll{11}}}\ ,
\xxn
\pp\sqd\phc = - \hh \cc{{\inv\ll{31}}\over{\inv\ll{11} }} + \hh \cc {{\inv\ll{21}\sqd}\over{\inv\ll{11}\sqd}}\ ,
\eee
and
\bbb
\pp\pb \phc = - \hh \cc {{\inv\ll{22}}\over{\inv\ll{11}}} + \hh \cc  {{\inv\ll{21} \inv\ll{12}}\over{\inv\ll{11}\sqd}}\ ,
\eee
such that
\bbb
(\pp\pb \phc ) \big ({{ (\pp X)\sqd }\over{ \pp X \cdot \pb X}} \big )
= (\pp X)\sqd \cc\left(  - \hh \cc
{{\inv\ll{22}}\over{\inv\ll{11}\sqd}} 
 + \hh \cc  {{\inv\ll{21} \inv\ll{12}}\over{\inv\ll{11}\uu 3}}  \right)\ .
\eee

\subsection{EOM and stress tensor conservation}

Next we will write the EOM for the $X\uu\m$ coordinates and
verify the classical conservation of the stress tensor,
when the EOM is satisfied.

\subsubsection{Deriving the EOM}

Under a general variation of $X$, $X\uu\m \to X\uu\m + \e\uu\m$, the field $\phc$ varies as
\bbb
\d \phc = - {{\gg X \cdot\gg  \e}\over{(\gg X)\sqd}}\ .
\eee
The variation of the action is
\be
0 &=& - \pi \apr \d S\ll{{\rm E}} =
 \int \cc d\sqd \s \cc \sqrt{|g|} \cc \e\ll\m\cc \biggl \{ \hh\cc \gg\sqd X\uu\m
+ \b\apr \cc \gg\uu a \bigg [ {{\gg\ll a X\uu\m }\over{(\gg X)\sqd}}\cc \gg\sqd \phc \bigg ] \biggl \} 
\nn\\
&& + \, ({\rm terms~involving~}{\cal R}\lp 2) \ ,
\ee
so the EOM is
\bbb
\gg\sqd \cc X\uu\m = - 2 \b\apr \cc  \gg\uu a \bigg [ {{\gg\ll a X\uu\m }\over{(\gg X)\sqd}}\cc \gg\sqd \phc \bigg ]
+ ({\rm terms~involving~}{\cal R}\lp 2)\ .
\eee
On flat space we have
\bbb
\vec{\pp}\sqd \cc X\uu\m = - 2 \b\apr \cc  \pp\uu a \bigg [ {{\pp\ll a X\uu\m }\over{(\vec{\pp} X)\sqd}}\cc \vec{\pp}\sqd \phc \bigg ]\ ,
\eee
where $\vec{\pp}\sqd$ is the flat Laplacian $\pp\ll b \pp\ll b$.  Written in terms of 
$w$ and $\wb$, we have
\bbb
\pp\pb X\uu\m = -  \b\apr \cc  \pp \bigg [ {{\pb X\uu\m }\over{\inv\ll{11}}}\cc \pp\pb \phc \bigg ]
 -  \b\apr \cc  \pb \bigg [ {{\pp X\uu\m }\over{\inv\ll{11}}}\cc \pp\pb \phc \bigg ]\ .
\een{EqnAbove}

It is helpful to recall the usual logic about why we can discard terms proportional to $\pp\pb X$ on
the right-hand side, if we are only working to order $\b \cc |X|\uu{-2}$ relative to leading-order quantities.
The logic is that our EOM are of the form $\pp\pb X\uu\m = O( \b / |X|)$.  Now
the order $\b / |X|$ terms on the right-hand side can be separated into terms without $\pp\pb X$ factors,
and terms with $\pp\pb X$ factors.  Both are at most of order $\b / |X|$, of course, but the latter
is equal to $(\pp\pb X)\uu\n \cc S\ll\n$, where $S\ll \n$ is of order $\b / |X|\sqd$.  However, $\pp\pb X$ is
itself of order $\b / |X|$.  We then have
\bbb
\pp\pb X\uu\m = \big \{ {\rm terms~obtained~by~discarding~}\pp\pb X ~{\rm on~the~RHS~of~}\rr{EqnAbove} \big \}
+ O(\b\sqd / |X|\uu 3)\ .
\eee
Letting the equivalence $\simeq$ denote equality up to terms of $O(\b\sqd / |X|\uu 3)$, we obtain
\bbb
\pp\pb X\uu\m \simeq 
 - \b\apr \cc \bigg (   \pb X\uu\m\cc  \pp \bigg [ { 1\over{\inv\ll{11}}}\cc \pp\pb \phc \bigg ]
+  \pp X\uu\m \cc  \pb \bigg [ {1 \over{\inv\ll{11}}}\cc \pp\pb \phc \bigg ]
    \bigg ) \ .
\een{EOMInConformallyNaturalXFields}

\subsubsection{Verifying the holomorphy of the stress tensor}

To evaluate the nonconservation of the order-$\b\uu 0$ holomorphic stress tensor, we can contract \rr{EOMInConformallyNaturalXFields} with 
$- {2\over{\apr}} \cc\pp X\ll\m$, to obtain:
\bbb
\pb T\ups{\b\uu 0} = - {1\over{\apr}}\pb (\pp X\cdot\pp X) 
=  - {2\over{\apr}} \cc \pp X\ll\m\cdot \pp\pb X\uu\m
\xxn
\simeq  2\b\pp X\ll\m\cc\biggl \{ \cc \pb \bigg [ {{\pp\pb \phc}\over{\inv\ll{11}}}\pp X\uu\m \bigg ] \cc 
+ \cc \pp \bigg [ {{\pp\pb \phc}\over{\inv\ll{11}}}\bp X\uu\m \bigg ] \biggl \}\ 
\xxn
 = 2\b\biggl\{
{{\bp(\pp X\cdot\pp X)}\over{\inv\ll{11}}}\pp\pb\phc+\pp^2\pb\phc-{{\pp {\inv\ll{11}}}\over{{\inv\ll{11}}}}\pp\pb\phc
+\frac{\pp X\cdot\pp X}{{\inv\ll{11}}}\pp\pb^2\phc-\frac{\pp X\cdot\pp X}{{\inv\ll{11}}^2}\pb{\inv\ll{11}}\pp\pb\phc
\biggl\}.
\eee
Now we can also evaluate the nonconservation of the order-$\b\uu 1$ piece of the stress tensor. 
Using ${{26-D}\over 6} = +2\b$, we have
\bbb
\pb T\ups{\b\uu 1} = -2\b\cc \pb \cc \biggl \{ (\pp\phc)\sqd +  {{(\pp X)\sqd \cc \pp\pb \phc}\over{\inv\ll{11}}}
 + \pp\sqd\phc \biggl \}
\xxn
=-2\b\biggl\{
2\p\phc\p\pb\phc+\frac{\pb(\p X\cdot\p X)}{{\inv\ll{11}}}\p\pb\phc+\frac{\p X\cdot\p X}{{\inv\ll{11}}}\p\pb^2\phc
-\p X\cdot\p X\p\pb\phc\frac{\pb {\inv\ll{11}}}{{\inv\ll{11}}^2}+\p^2\pb\phc
\biggl\}.
\eee

These two nonconservations cancel each other out exactly, so we conclude:
\bbb
\pb T\ups{\b\uu 0} + \pb T\ups{\b\uu 1} \simeq 0  \ , 
\eee
where the $\simeq$ denotes equality up to terms of order $\b\sqd / |X|\sqd$.

\subsection{Relation to the old covariant formalism}
We now relate our embedding of the effective string in the Polyakov formalism to the old
Lorentz-covariant formalism of \cite{Polchinski:1991ax}.  Apart from the addition of the intrinsic metric and the
corresponding Weyl invariance, the relationship between the two actions also includes  
a redefinition of the $X$ variables at next-to-leading order.
 
Our stress tensor is not conserved if we apply the EOM derived from
the Polchinski-Strominger action in the form written in \cite{Polchinski:1991ax}.  This has
to do with the fact that our composite-Liouville action differs from the PS action by 
linear combinations of $\pp\pb X\uu\m$ multiplied by operators, meaning that the $X$-variables here differ from those in 
\cite{Polchinski:1991ax} by a field transformation of the form
$X\uu\m \to X\uu\m + {\cal O}\uu\m$, where 
\bbb
{\cal O^\mu}=\frac{\b\alpha^\prime}{4}\frac{1}{\inv_{11}^2}\biggl\{ (\inv_{21}+\partial X\cdot\partial\bar\partial X)\bar{\partial}X^\mu+\inv_{12} \partial X^\mu\biggl\}\ .
\een{UsToPSFieldRedef}
To derive this, we first find terms of the 
form $F[X]^\mu\partial\bar\partial X_\mu$ in our composite-Liouville action. 
Then, $F[X]^\mu$ and $\mathcal O^\mu$ are related by ${\mathcal O}^\mu=-\pi\alpha^\prime F[X]^\mu$.

The useful aspect of this change of variables has to 
do with the conformal properties of the $X\uu\m$ field.  The worldsheet coordinate $X\uu\m\ll{\textsc{PS}}$
as defined in \cite{Polchinski:1991ax} does not have simple conformal properties.  The operator product expansion between the stress tensor
and $X\uu\m\ll{\textsc{PS}}$ has a singular term $z\uu{-3} \cc \b\cc {{\pb X\uu\m}\over{\inv\ll{11}}}$, so the coordinate $X\uu\m\ll{\textsc{PS}}$ is not
a conformal primary.  We shall see in section \rr{TXOPESec} that our embedding coordinate 
$X\uu\m\ll{{\rm{here}}}$ has leading term
$z\uu{-1} \cc \pp X\uu\m\ll{{\rm{here}}}$ in its OPE with the stress tensor.  The $Xll{{\rm{here}}}$ variables in our formalism
are more natural objects from a conformal point of view, which follows from our derivation of the composite-Liouville interaction term from a Weyl-invariant
path integral in which $X\uu\m\ll{{\rm{here}}}$ is taken 
to be invariant under Weyl rescalings.\footnote{The same observation was made
in \cite{Hari Dass:2007gn}.}  We shall establish the quantum conformal properties of our
embedding coordinates $X\uu\m\ll{{\rm{here}}}$ by explicit calculation in section \ref{PropCorr}.

\section{Local operators in covariant effective string theory}

Let us now classify Weyl-invariant operators, organizing them by their $X$-scaling.  We begin by constructing
a Weyl tensor calculus based on a Weyl-covariant derivative using the composite Liouville field $\phc$.
(Essentially the same Weyl tensor calculus was developed in \cite{Hari Dass:2007gn}.)
\subsection{Weyl tensor calculus on the worldsheet }

The ordinary Riemannian covariant derivative does not transform covariantly under Weyl transformations.  The
Riemannian connection transforms as
\bbb
\d (\gg\ll b V\uu a) =(\pp\ll b \o) V\uu a + \d\uu a{}\ll b (V\uu c\pp\ll c\o) - (\pp\uu a\o) V\ll b 
\eee
under a Weyl transformation parametrized by $\o$, if $V$ itself is neutral under Weyl rescalings.  However, using
the effective Liouville field $\phc$, which simply shifts under a Weyl transformation \rr{PhiCompTrans},
we can easily render our Riemannian connection Weyl-covariant.
Defining the Weyl-covariant derivative
\bbb
\hat{\gg}\ll b V\uu a \equiv \gg\ll b V\uu a - (\pp\ll b \phc) V\uu a - \d\uu a{}\ll b (V\uu c\pp\ll c\phc) +
(\pp\uu a\phc) V\ll b\ ,
\eee
$\hat{\gg}\ll b V\uu a$ transforms as a tensor of type $(1,1)$ under diffeomorphisms, and trivially
under Weyl transformations:
\bbb
\d\lp{{\rm Weyl}} \hat{\gg}\ll b V\uu a = 0\ ,
\eee
if we assume the property \rr{PhiCompTrans}.
The action of our diff- and Weyl-covariant derivative on covectors is
\bbb
\hat{\gg}\ll a W\ll b \equiv \gg\ll a W\ll b 
 + (\pp\ll b \phc) W\ll a + (\pp\ll a \phc) W\ll b - g\ll{ab} (W\ll c\pp\uu c \phc)\ ,
\eee
if $W$ transforms trivially under Weyl transformations.
So, in particular, if $W\ll b \equiv \hat{\gg}\ll b Y =  \gg\ll b Y = \pp\ll b Y$ for some scalar $Y$, then
\bbb
\hat{\gg}\ll a \hat{\gg}\ll b Y =\hat{\gg}\ll a \gg\ll b Y = \gg\ll a \gg\ll b Y + \pp\ll a \phc \pp\ll b Y
+ \pp\ll b \phc \pp\ll a Y -  g\ll{ab} \pp\uu c \phc \pp\ll c Y\ .
\eee

We conclude by commenting on this Weyl-covariant double derivative.  First,
it is automatically symmetric when acting on scalars, just like the Riemannian covariant derivative:
\bbb
\hat{\gg}\ll a \hat{\gg}\ll b Y = \hat{\gg}\ll b \hat{\gg}\ll a Y\ .
\een{WeylCovDerSym}
Second, the correction terms implementing the Weyl-covariance are traceless, so
the Weyl-covariant Laplacian is equal to the ordinary Riemannian Laplacian:
\bbb
\hat{\gg}\uu a \hat{\gg}\ll a Y = \gg\uu a \gg\ll a Y\ .
\een{WeylCovDerTraceless}
These identities will prove themselves convenient in the calculations that follow.

\subsection{Construction of Weyl tensor local operators}

Now we can construct diffeomorphism-invariants with definite Weyl scaling, 
which will comprise the basic ingredients of the gauge-invariant
operators we will discuss later on.  
Starting with an arbitrary scalar $Y$, we can construct a four-derivative invariant by taking the norm-squared of the tensor \rr{WeylCovDerTraceless}, 
\be
\co &\equiv& g\uu{ac} g\uu{bd} \cc \hat{\gg}\ll a \hat{\gg}\ll b Y \cc  \hat{\gg}\ll c \hat{\gg}\ll d Y
\nn\\
&=& (\gg\ll a \gg\ll b Y)(\gg\uu a\gg\uu b Y) + 4 \cc (\pp\ll a \phc)(\pp\ll b Y)(\gg\uu a\gg\uu b Y) 
-  2\cc (\pp\ll a \phc)(\pp\uu a Y)(\gg\sqd Y) 
\nn\\
&& + 4\cc (\vec{\pp} \phc)\sqd (\vec{\pp} Y)\sqd - 2\cc (\pp\ll a\phc \pp\uu a Y)\sqd\ .
\ee
This operator $\co$ is a scalar under diffeomorphisms, and transforms with weight $-4$ under
Weyl transformations, in units where $g\ll{\bullet\bullet}$ has weight $+2$.

Now we consider the case $Y = X\uu\m$, transforming
the expression to unit gauge.  We begin by defining
\bbb
\ca\ll{(ab)(cd)} \equiv \gg\ll a \gg\ll b X \cdot \gg\ll c \gg\ll d X\ .
\eee
Now we would like to find the unit-gauge expressions for 
\bbb
\hat{\ca}\ll{(ab)(cd)} \equiv \hat{\gg}\ll a \hat{\gg}\ll b X \cdot \hat{\gg}\ll c \hat{\gg}\ll d X\ ,
\eee
and
\bbb
\co\ll{22} \equiv g\uu{ac} \cc g\uu{bd} \cc  \hat{\ca}\ll{(ab)(cd)}\ .
\eee
We start by considering just the values for the second derivative, using symmetries, 
the leading-order EOM, and Virasoro constraints.
We obtain
\bbb
\hat{\gg}\ll b \hat{\gg}\ll a X\uu\m = \hat{\gg}\ll b \pp\ll a X\uu\m = \gg\ll b\pp\ll a X\uu\m
+ \pp\ll b \phc \cc \gg\ll a X\uu\m +  \pp\ll a \phc \cc \gg\ll b X\uu\m - g\ll{ab} \cc \pp\uu c \phc \cc \gg\ll c X\uu\m\ ,
\eee 
with the properties
\bbb
\hat{\gg}\ll b \hat{\gg}\ll a X\uu\m = \hat{\gg}\ll a \hat{\gg}\ll b X\uu\m \ ,
\eee
and
\bbb
g\uu{ab} \cc \hat{\gg}\ll b \hat{\gg}\ll a X\uu\m = g\uu{ab} \cc \gg\ll b\gg\ll a X\uu\m \simeq 0\ ,
\eee
where the $\simeq$ means we are modding out by the leading-order EOM.

In unit gauge, where $g\ll{ab} = \d\ll{ab}$, we have
\bbb
\hat{\gg}\ll w \hat{\gg}\ll \wb X\uu\m = \hat{\gg}\ll \wb \hat{\gg}\ll w X\uu\m \simeq 0\ .
\eee
The other components are
\bbb
\hat{\gg}\ll w\sqd X\uu\m = \pp\ll w\sqd X\uu\m + 2 \cc\pp\ll w \phc \cc \pp\ll w X\uu\m \simeq
 \pp\ll w\sqd X\uu\m - {{\pp\ll w\sqd X \cdot \pp\ll\wb X}\over{\pp\ll w X \cdot \pp\ll\wb X}} \cc \pp\ll w X\uu\m
 = \pp\ll w\sqd X\uu\m - {{\inv\ll{21}}\over{\inv\ll{11}}} \cc \pp\ll w X\uu\m\ ,
\xxn
\hat{\gg}\ll \wb\sqd X\uu\m = \pp\ll \wb\sqd X\uu\m + 2 \cc\pp\ll \wb \phc \cc \pp\ll \wb X\uu\m
\simeq
 \pp\ll \wb\sqd X\uu\m - {{\pp\ll w X \cdot \pp\ll \wb\sqd X  }\over{\pp\ll w X \cdot \pp\ll\wb X}} \cc \pp\ll \wb X\uu\m
 = \pp\ll \wb\sqd X\uu\m - {{\inv\ll{12}}\over{\inv\ll{11}}} \cc \pp\ll \wb X\uu\m\ ,
\eee
so that
\be
\widehat{\inv}\ll{22} &\equiv& \hat{\gg}\ll w\sqd X\uu\m \cc \hat{\gg}\ll \wb\sqd X\ll\m 
\nn\\
&=& \big ( \cc   \pp\ll w\sqd X\uu\m - {{\inv\ll{21}}\over{\inv\ll{11}}} \cc \pp\ll w X\uu\m   \cc \big ) \cc   
\big ( \cc  \pp\ll \wb\sqd X\ll\m - {{\inv\ll{12}}\over{\inv\ll{11}}} \cc \pp\ll \wb X\ll\m  \cc \big )  
\nn\\
&=& \inv\ll{22} - {{\inv\ll{12} \inv\ll{21}}\over{\inv\ll{11}}}\ .
\ee

Likewise, in Lorentzian signature, we have
\bbb
\widehat{\inv}\ll{22} \equiv \hat{\gg}\ll +\sqd X\uu\m \cc \hat{\gg}\ll 
- \sqd X\ll\m = \inv\ll{22} - {{\inv\ll{12} \inv\ll{21}}\over{\inv\ll{11}}}\ .
\een{WeylCovarX22}
The expression for this invariant in unit gauge can also be written as 
\bbb
\widehat{\inv}\ll{22} = \inv\ll{11} \pp\pb [{\rm ln}(\inv\ll{11})]\ .
\een{ConformalGaugeExpressionForHatInv22}
We will use the invariant $\widehat{\inv}\ll{22}$ in subsequent 
discussions to construct higher-dimension operators that will play a role as
terms in the action with adjustable coefficients.

The operator $\widehat{\inv}\ll{22}$, and functions of it and $\inv\ll{11}$, are the only Weyl-invariant scalar operators that
can be constructed using two or fewer derivatives.   
There is another candidate operator with only single- and double-derivatives acting on $X\uu\m$, but
we will see now that this operator vanishes.  That is,
\bbb
\co\upp{{\rm six-derivative}} \equiv (\hat{\gg}\ll a X \cdot \hat{\gg}\ll b \hat{\gg}\ll c X) \cc (\hat{\gg}\uu a X \cdot \hat{\gg}\uu b \hat{\gg}\uu c X)
\eee
vanishes.  This fact is easily understood in conformal gauge in 
the basis of the non-Weyl-covariant $\inv\ll{pq}$.  The difference between any covariant $\widehat{\inv}\ll{pq}$
and its non-covariant version $\inv\ll{pq}$ involves a set of correction terms that can be 
written in terms of $\inv\ll{rs}$, with $r < p$ or $s < q$, or both.  
The difference between the Weyl-covariant and non-Weyl-covariant versions of $\inv\ll{12} \inv\ll{21}$ 
would have to contain only $\inv\ll{11}\uu 3$.  
Terms with a single $\inv\ll{12}$ or a single $\inv\ll{21}$, 
but not both, would have to have spin $\pm 1$ and could not be a scalar.  
So the the operator $\co\upp{{\rm six-derivative}}$ can only be proportional to a sum of
$\inv\ll{12}\inv\ll{21}$ and $\inv\ll{11}\uu 3$.  Neither of these can appear.  The term $\inv\ll{11}\uu 3$ cannot
appear because the Weyl-covariant derivative preserves the $X$-scaling of an 
operator.  So $\inv\ll{11}\uu 3$, which has $X$-scaling $+6$, cannot appear in a covariant version 
of $\inv\ll{12}\inv\ll{21}$, which has $X$-scaling $+4$.  The noncovariant $\inv\ll{12} \inv\ll{21}$ has the
correct weight under rigid rescalings, but does not transform as a tensor under general conformal transformations,
as it is proportional to $\inv\ll{11}\sqd\cc (\pp \cc{\rm ln}(\inv\ll{11})) (\pb \cc {\rm ln}(\inv\ll{11}))$.  
Indeed, the noncovariance of $\inv\ll{12}\inv\ll{21}$ under Weyl transformations is the 
key property that allows it to appear as the numerator of the classical
term cancelling the quantum-mechanical Weyl anomaly.

\subsection{Higher-dimension operators in the string effective action}

\subsubsection{Conformal dimension and $X$-scaling}

In conformal gauge, all operators suitable for addition to the action are of course dimension $2$.  This is logically
necessary because conformal invariance is a residual gauge symmetry of the action.  However, effective field theory
is still applicable in the sense that operators are arranged hierarchically in terms of their $X$-scaling, with a maximum
$X$-scaling appearing ($X$-scaling $+2$), and with the relevance of operators in the long-string approximation determined by
their $X$-scalings.

Assuming the basic symmetries of the problem -- the target-space Poincar\'e symmetry and the worldsheet gauge symmetries --
we can classify all operators up to a given $X$-scaling, modulo total derivatives and terms proportional to the leading-order 
equations of motion and leading-order Virasoro constraints.  For straight, static strings, such as those 
considered in \cite{Luscher:2002qv, Luscher:2004ib, Aharony:2010db, Aharony:2011ga, Drummond:2004yp, Aharony:2009gg, Aharony:2010cx, Aharony:2011gb, Dubovsky:2012sh, AharonyUnPub, Dubovsky:2013gi, Aharony:2013ipa}, 
the computation of any amplitude to order $\apr{}\uu p / R\uu {2p}$ relative to its classical value
requires including all possible terms in the effective action
of order up to and including $|X|\uu{-2(p-1)}$.  For rotating strings, such as those considered in 
\cite{Baker:2000ci, Baker:2002km, Kruczenski:2004me, Sonnenschein:2014jwa, Hellerman:2013kba}, computation of amplitudes to order
$J\uu{-p}$, relative to the classical value, requires including the same set of operators.

\subsubsection{The $\inv\ll{11}$-dressing rule}

The problem is simplified by the hypothesis that the only quantity allowed to appear raised to
negative powers in local operators, is
the invariant $\inv\ll{11}$.  Treating this as an exact property of the effective string expansion, order by
order in the inverse size of the string, can be justified on several grounds.  
First, all other Poincar\'e-invariant combinations $\inv\ll{pq}$ vanish
in the static string configuration.  Therefore, bilinears $\inv\ll{pq}$ with $p$ or $q$ greater than or equal to $2$ cannot
appear as denominators 
in any effective action that can be used to describe a long, approximately static string.  Second, the
invariant $\inv\ll{11}$ is the operator with the largest ratio of $X$-scaling to worldsheet scaling dimension.  
Thus, any expression with homogeneous worldsheet scaling dimension that can be expanded as a sum of terms will be dominated at large
$X$ by the terms with the most powers of $\inv\ll{11}$.  For rational expressions, 
where both the numerator and denominator are polynomials
in multi-derivatives of $X$, the denominator in particular is dominated by terms with the largest number 
of powers of $\inv\ll{11}$.  Thus the rational function in question can be expanded at large $X$, where 
the denominator will always be dominated by the term
with the largest number of powers of $\inv\ll{11}$.   

For instance, the term $\co \equiv {{ \widehat{\inv\ll{33}}\sqd}\over{\inv\ll{11}\uu 5 + \widehat{\inv\ll{22}}\sqd \inv\ll{11}}}$
can be expanded at large $X$ to give
\bbb
\co = {{ \widehat{\inv\ll{33}}\sqd}\over{\inv\ll{11}\uu 5 }} - {{ \widehat{\inv\ll{33}}\sqd  \widehat{\inv\ll{22}}\sqd }\over{\inv\ll{11}\uu 9 }} + \cdots
\eee
with the omitted terms being of order $\widehat{\inv\ll{33}}\sqd  \widehat{\inv\ll{22}}\uu 4  / \inv\ll{11}\uu{13} = O(|X|\uu{-14})$.  Terms such
as $\widehat{\inv\ll{33}}\sqd / \widehat{\inv\ll{22}}\sqd$ are consistent with all the necessary symmetries, 
but they are singular in the long string vacuum and thus
describe a different universality class from the conventional effective string theory.  
They are also fine-tuned, in the sense that a small perturbation of the
microscopic theory could be expected to change the denominator 
from $\widehat{\inv\ll{22}}\sqd $ to $\widehat{\inv\ll{22}}\sqd + \e\pr \inv\ll{11} \inv\ll{33} + \e \inv\ll{11}\uu 4 $ for
some small $\e$ and $\e\pr$.   
These three terms have $X$-scaling $|X|\uu 4, |X|\uu 4,$ and $|X|\uu 8$, respectively, 
so no matter how small the value of $\e$, the denominator will always be dominated 
by the $\inv\ll{11}\uu 4$ term for string solutions with a sufficiently long scale $|X| \sim R$.  
For the theory to be dominated by $\widehat{\inv\ll{22}}\sqd$
in the long-string limit, the coefficients $\e, \e\pr$ would have to be fine-tuned exactly 
to zero, in the absence of any symmetry principle that would enforce this independently.  The
$\inv\ll{11}$ dressing rule is, in this sense, just a corollary of the principle of naturalness.

The third and most concrete justification for the $\inv\ll{11}$-dressing rule is that it appears to be true.   
We know of several examples of microscopically well-defined holographic theories that give rise to effective 
string theories in 4D, after integrating out the motions
of the string into the fifth (and higher) dimension.  In each of these examples, the dressing rule appears to hold.  
For instance, the effective strings generated by holographic compactifications \cite{Aharony:2013ipa} 
obey the rule, as does the effective theory generated by the irrelevantly-perturbed Liouville theory discussed 
at the end of \cite{Polchinski:1991ax}, and the effective theory in an early derivation the 4D effective string from
warped compactification with a minimum of the warp factor \cite{natsuume}.  
Though the analysis in these works was at the classical and one-loop level, it is clear that
the dressing rule holds in these models to all loops.  
The mass $M\ll{{\rm holo}}$ of the fluctuations into the holographic direction (the mass being
taken with respect to conformal time on the worldsheet) is proportional to $\inv\ll{11}$ 
at large $|X|$, so when one integrates out the holographic direction the singular operators 
derive from inverse powers of $M\ll{{\rm holo}}$, and thus have $\inv\ll{11}$ dressing.

The dressing rule must clearly break down near a point on the worldsheet where $\inv\ll{11}$ vanishes.  Such points appear on the worldsheets of any realistic rotating string solution in four dimensions, such as closed strings or open strings with Neumann boundary conditions.  In the vicinity of such points the $\inv\ll{11}$-dressing rule is violated, but the effective theory 
need not itself break down or include additional light degrees of freedom.  
In the case of strings with massless quarks, the singularity is resolved by a 
reorganization of operators dressed with
$\inv\ll{22}$, rather than $\inv\ll{11}$ \cite{Hellerman:2013kba}.  
This reorganization of operators is physically interesting and may be testable experimentally: The emergence of the
$\inv\ll{22}$ dressing is associated with fractional powers of angular momentum appearing in the formula for the large-$J$
meson spectrum \cite{Baker:2002km,Wilczek:2004im,Sonnenschein:2014jwa,Kruczenski:2004me}. We will
not deal with such situations in this paper, however; we restrict our attention to local properties of
the worldsheet effective theory away from loci where $\inv\ll{11}$ vanishes.

\subsubsection{Scalar operators of conformal dimension $2$}

\def\hinv{{\blue{{\cal H}}}}
\def\ainv{{\blue{\tilde{{\cal H}}}}}

Assuming the validity of the $\inv\ll{11}$-dressing rule,
the enumeration of all gauge-invariant operators at a given order in $|X|$ 
becomes a finite problem.  The set of Poincar\'e invariant terms
is generated by the bilinears $\inv\ll{pq}$, together with 
$\hinv\ll{(pq)} \equiv \pp\uu p X \cdot \pp\uu q X$ and $\ainv\ll{(pq)}\equiv \pb\uu p X \cdot \pb\uu q X$.  
These operators, and their
covariantized versions $\widehat{\inv\ll{pq}}, \widehat{\hinv\ll{(pq)}}, \widehat{\ainv\ll{(pq)}} $, 
have scaling dimension $p+q$.  To classify possible
terms in the action, we can write each operator uniquely as an 
undressed numerator (i.e., a monomial or polynomial in $\hinv\lp{rs}, \ainv\lp{tu}$, and $\inv\ll{pq}$ 
with $p$ or $q \geq 2$) dressed with a power of $\inv\ll{11}$.  We may then discard all terms proportional to leading-order Virasoro 
constraints and total derivatives.\footnote{We have chosen our basis of operators so
that terms proportional to the leading-order EOM $\pp\pb X$ are already omitted.}

The power of $\inv\ll{11}$ dressing a given undressed operator is dictated uniquely by scale invariance, with an underssed scalar operator 
\bbb
\co\uprm{undressed} \equiv \prod\ll i \inv\ll{p\ll i q\ll i} \cc\prod\ll j  \hinv\lp{r\ll j s\ll j} 
\cc\prod\ll k  \ainv\lp{t\ll k u\ll k}\ , 
\eee
dressed to marginality as
\bbb
\co\uprm{dressed} \equiv \inv\ll{11}\uu{-\D} \co\uprm{undressed}\ ,
\xxn
\D = -1 + \sum\ll i p\ll i + \sum\ll j \big [ \cc r\ll j + s\ll j \cc \big ]  = -1 + \sum\ll i q\ll i + \sum\ll k \big [ \cc t\ll k + u\ll k \cc \big ] \ .
\eee
The equality of the latter two expressions is simply the constraint that the operator be a scalar field.  The $X$-scaling of the undressed operator is
$2(N\ll\inv + N\ll\hinv + N\ll\ainv) = \sum\ll i 2 + \sum\ll j 2 + \sum\ll k 2$,
 so the $X$-scaling of the dressed operator is
\be
{\tt Scale\ll X}\big [ \co\uprm{dressed} \big ] &=& -2 \D + {\tt {\tt Scale\ll X}}\big [ \co\uprm{undressed} \big ] 
\nn\\
&& \kern-35pt = 2 - \sum\ll i 2 (p\ll i - 1) - \sum\ll j \big [ \cc 2(r\ll j - 1) + 2 (s\ll j - 1) \cc \big ]
\nn\\
&& \kern-35pt = 2 - \sum\ll i 2 (q\ll i - 1) - \sum\ll k \big [ \cc 2(t\ll k - 1) + 2 (u\ll k - 1) \cc \big ] \ .
\ee
We can write this symmetrically as
\be
{\tt Scale\ll X}\big [ \co\uprm{dressed} \big ] &=&
 2 - \sum\ll i \big [ \cc (p\ll i - 1) + (q\ll i - 1) \cc \big ]  
\nn\\
&&\kern-55pt - \sum\ll j \big [ \cc (r\ll i - 1) + (s\ll i - 1) \cc \big ] -
\sum\ll k  \cc \big [ (t\ll k - 1) + (u\ll k - 1)  \cc \big ] \ .
\label{XScalingFormulaSymmetric}\ee
The invariants $\hinv\lp{11}$ and $\ainv\lp{11}$ are proportional to the leading-order Virasoro constraints, so every $\inv, \hinv$ or 
$\ainv$ added to an operator contributes negatively to the $X$-scaling, from the starting point of the tree-level action $\inv\ll{11}$, 
with $X$-scaling $2$.

\subsection{Universality and nonuniversality}\label{UnivAndNonuniv}
In this section we will analyze the issue of universality of the action at the first two
subleading orders in the $1/R$ expansion, where $R$ is the typical length scale of the string.  
We will emphasize that all amplitudes are universal at next-to-leading
order (NLO).  At next-to-next-to-leading order (NNLO), it will emerge that certain observables are universal
and some are not, and we will explain why this is so, giving a useful criterion to
distinguish between the two cases.

\subsubsection{Universality at NLO}
To classify possible marginal operators with $X$-scaling $0$, we must, according to \rr{XScalingFormulaSymmetric}, 
find all undressed operators with
\bbb
 \sum\ll i \big [ \cc (p\ll i - 1) + (q\ll i - 1) \cc \big ]  + \sum\ll j \big [ \cc (r\ll i - 1) + (s\ll i - 1) \cc \big ] + \sum\ll k  \cc \big [ (t\ll k - 1) + (u\ll k - 1)  \cc \big ]  = 2\ .
\eee
The only such operators are $\hinv\lp{12} \ainv\lp{12}$, $\inv\ll{12}\inv\ll{21}$ and $\widehat{\inv\ll{22}}$.  The first vanishes, modulo operators with lower $X$-scaling,
by the Virasoro constraints.  The second, as discussed above, 
does not transform covariantly under Weyl transformations, cannot be rendered covariant with
additional terms, and does not correspond to a gauge-invariant operator after dressing with 
two negative powers of $\inv\ll{11}$.  This leaves us with 
the Weyl-covariant scalar $\widehat{\inv\ll{22}}$, which is dressed to 
marginality as $\widehat{\inv\ll{22}} / \inv\ll{11}$.  From the form \rr{ConformalGaugeExpressionForHatInv22},
we see this operator is proportional to a total derivative, and does not affect the string 
dynamics except possibly as a boundary term.  This leaves no adjustable operators 
in the string action at next-to-leading order in the large $X$ expansion for closed strings.  
This argument justifies the claim of universality
of the asymptotic Regge intercept for closed strings with angular momentum in two planes 
in $D\geq 5$.\footnote{These arguments do not apply without modification to the case of
strings with folds or Neumann boundaries, in which cases the dressing hypothesis 
breaks down.  Note, however, that in \cite{Hellerman:2013kba} a similar analysis was applied to derive
a universal asymptotic intercept for open string Regge trajectories under the hypothesis 
of an $\inv\ll{22}$ dressing rule for Neumann boundaries.}

\subsubsection{Nonuniversality at NNLO}
A similar analysis can be applied to classify operators at NNLO.
Here, we need to find undressed operators with 
\bbb
\sum\ll i \big [ \cc (p\ll i - 1) + (q\ll i - 1) \cc \big ]  + \sum\ll j \big [ \cc (r\ll i - 1) + (s\ll i - 1) \cc \big ] + \sum\ll k  \cc \big [ (t\ll k - 1) + (u\ll k - 1)  \cc \big ]  = 4\ .
\eee
There are many such operators, for instance:
\bbb
\co\uprm{undressed}\ll 1 \equiv (\widehat{\inv\ll{22}})\sqd, 
\xxn
\co\uprm{undressed}\ll 2 \equiv \widehat{\inv\ll{22}} \inv\ll{12} \inv\ll{21} \ , \llsk\llsk  \co\uprm{undressed}\ll 3 \equiv (\inv\ll{12}\inv\ll{21})\sqd\ ,
\xxn
\co\uprm{undressed}\ll 4 \equiv \widehat{\inv\ll{22}} \hinv\lp{12} \ainv\lp{12}\ ,  \llsk\llsk \co\uprm{undressed}\ll 5 \equiv      \hinv\lp{12}\sqd \ainv\lp{12}\sqd\ .
\eee
Most of these can be eliminated as candidate terms in the action.  Operators $\co\uprm{undressed}\ll 2$ and
$\co\uprm{undressed}\ll 3$ contain the expression $\inv\ll{12}\inv\ll{21}$, 
which is noncovariant, and cannot be rendered covariant, and
no linear combination of $\co\uprm{undressed}\ll 2$ and
$\co\uprm{undressed}\ll 3$ is covariant either.  The operators $\co\uprm{undressed}\ll 4$ and
$\co\uprm{undressed}\ll 5$ are proportional to the first 
derivatives of free stress tensors, and so vanish by the 
Virasoro constraints, modulo operators with smaller $X$-scaling.
The operator $\co\uprm{dressed}\ll 1 \equiv (\widehat{\inv\ll{22}})\sqd / \inv\ll{11}\uu 3$ is 
nonzero and independent as a possible term in the effective Lagrangian, however.
In gauge-invariant language, this corresponds to the curvature-squared of the induced metric.  
As a result, the predictions of effective string theory are nonuniversal at NNLO
in the long string expansion. 

The end of universality does not mean the end of the usefulness of effective string theory, however.  
The small number of adjustable terms means that there are far more observables than parameters at NNLO, so one
can derive long-string or large-$J$ sum rules that are universal at NNLO, because 
the adjustable coefficients cancel out in certain combinations of observables.
Also, in many situations, such as that of a static string, the term $\co\uprm{dressed}\ll 1 $ 
does not contribute at NNLO because its classical value vanishes.  Therefore, it
can only contribute through quantum fluctuations, which are suppressed by additional powers of $\apr / R\sqd$.  
As a result, the energy spectrum of the static string is universal
up to and including order $\apr R\uu{-3}$, rather than just $R\uu{-1}$.  
This NNLO universality holds only for the static string and other cases where $\co\uprm{dressed}\ll 1 $
vanishes classically.  It does not apply, for instance, to the case of the rotating string, 
so we should expect the meson mass spectrum to be sensitive to the coefficient of the
curvature-squared term at order $J\uu{-{3\over 2}}$.

\newpage
\section{Interaction corrections to the OPE}\label{PropCorr}

\newcommand{\PicFeynA}{  $\hbox{ \includegraphics[width=3.0in,height=4.5in,angle=0]{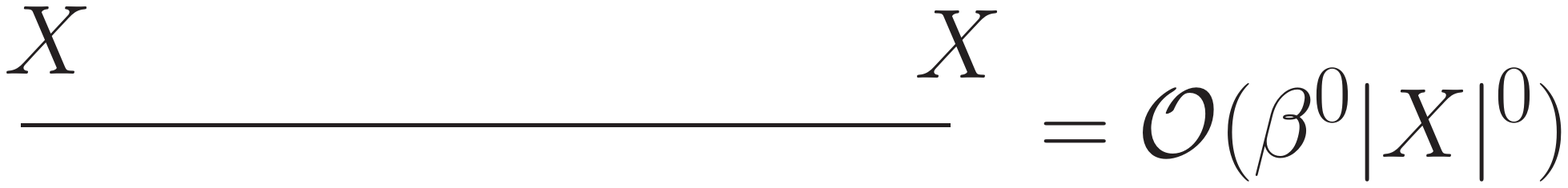}}$ }
\newcommand{\PicFeynB}{  $\hbox{ \includegraphics[width=3.63in,height=5.45in,angle=0]{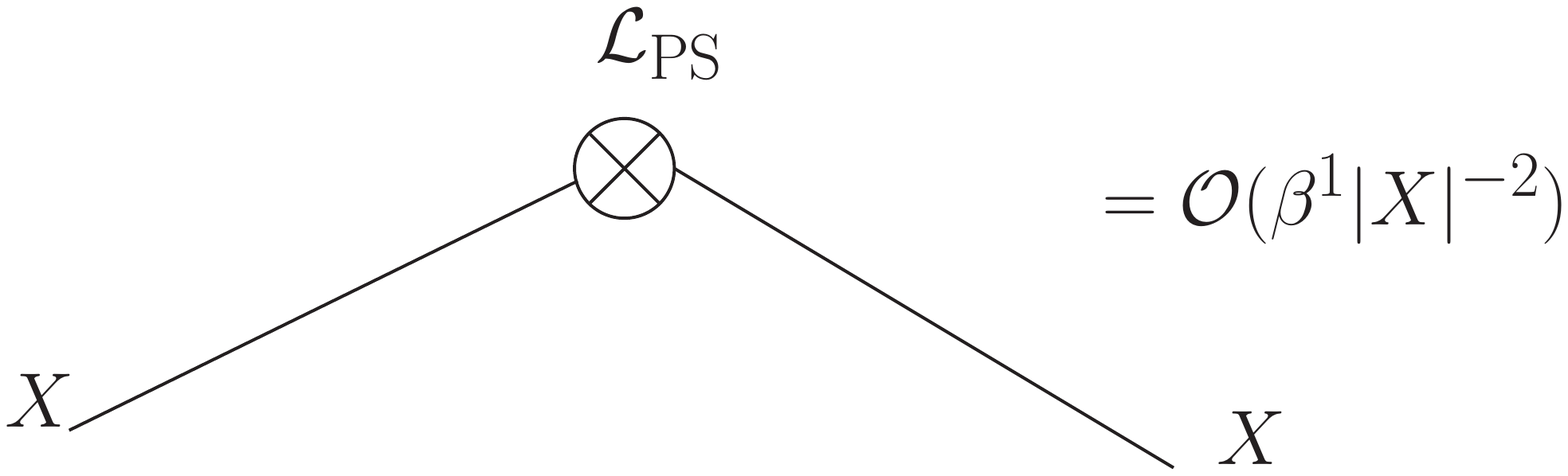}}$ }
\newcommand{\PicFeynC}{  $\hbox{ \includegraphics[width=3.15in,height=4.72in,angle=0]{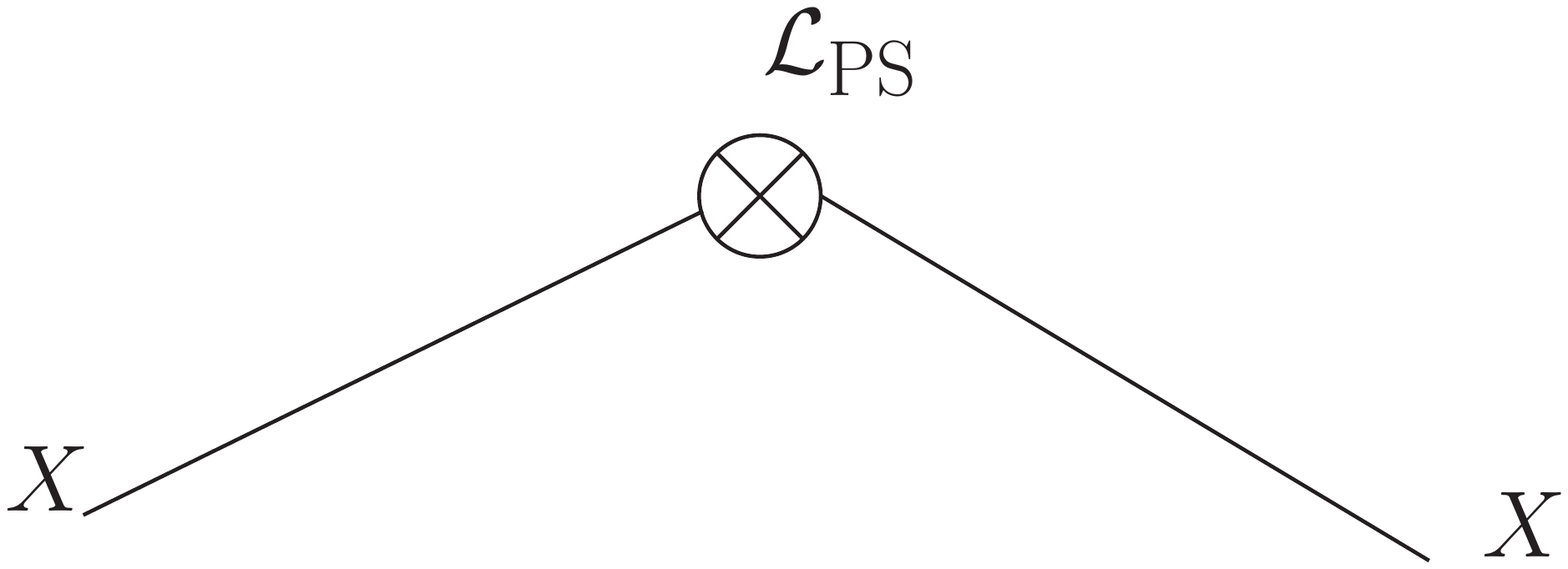}}$ }
\newcommand{\PicFeynD}{  $\hbox{ \includegraphics[width=1.1in,height=1.65in,angle=0]{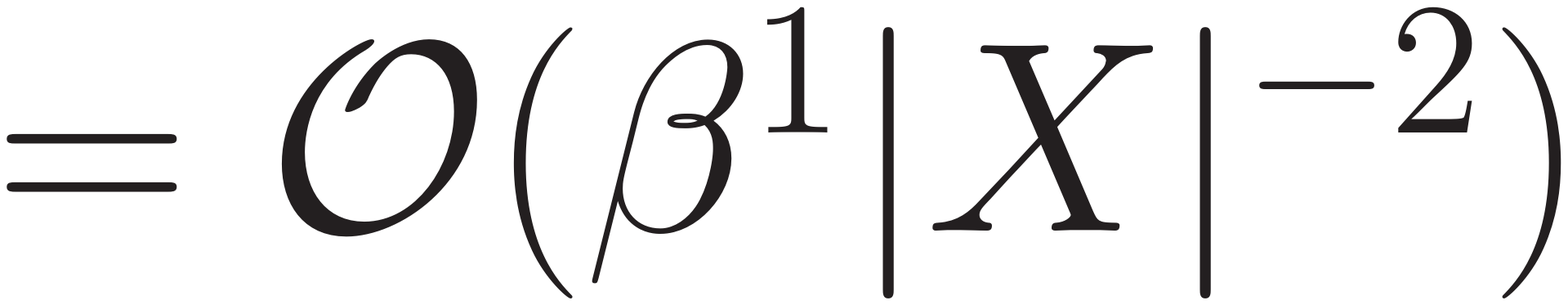}}$ }

\putat{115}{-190}{\PicFeynA}
\putat{70}{-260}{\PicFeynC}
\putat{247}{-140}{\PicFeynD}

\vskip2in
\subsection{The propagator correction}

In the presence of the interaction term, the theory preserves conformal invariance  
but is no longer a free conformal field theory.
The structure of the operator product expansion becomes less simple when $X$ is no longer a free field.  
The propagator for the fluctuation $Y\uu\m = X\uu\m - E\uu\m$ of the embedding coordinates 
receives non-vanishing corrections at relative order $\b |X|\uu{-2}$,
and these corrections affect the operator product expansions of local operators with one another.

The form of the leading-order propagator correction is generally not simple.
The correction satisfies a differential equation controlled by the $\b\uu 1 |Y|\uu 1$ terms in the expansion of $X$ in the EOM \rr{EOMInConformallyNaturalXFields} as
the solution $X\uu\m = E\uu\m + Y\uu\m$.  The full set of terms in the EOM 
for the propagator correction is lengthy, and we do not reproduce it here.  Rather,
we will deal directly with the effect of the propagator correction on the OPE of
the stress tensor with other operators.

Note that the propagator correction vanishes at first subleading order in the case of 
the static string \cite{Luscher:2002qv,Luscher:2004ib,Aharony:2010db,Aharony:2011ga,Drummond:2004yp,Aharony:2009gg,Aharony:2010cx,Aharony:2011gb,Dubovsky:2012sh,AharonyUnPub,Dubovsky:2013gi,Aharony:2013ipa}.  The corrections
to the tree-level propagator are all proportional to 
\bbb
E\ll{pq} \equiv  \inv\ll{pq} \big |\ll{X= E} = \pp\uu p E\cdot\pb\uu q E\ ,
\een{EpqDef} 
with $p$ and/or $q$ greater than unity, and
all such expectation values of higher-derivative invariants vanish in the classical solution for the static case.  
This is not so for more generic situations, such as the rotating string.  
The propagator is used to compute vertex operator correlation functions, as well as to verify the
consistency of the gauge-fixed theory via the stress tensor OPE.  
We therefore need to understand the propagator correction to some extent.


\subsection{Nonrenormalization of the OPE between $T$ and $X$ at $O(\b\uu 1)$}\label{TXOPESec}

 We will
see that the propagator correction, despite modifying the operator product expansion of the $X$ field, does not affect its conformal properties.  This
nontrivial fact, which is not manifest in the old covariant formalism of \cite{Polchinski:1991ax}, can be traced to our embedding of the effective string
in the Polyakov path integral.

In this subsection, the OPE of $T$ with $X$ is examined. Although there are the propagator corrections and the contribution from $T^{[\beta^1]}$, it will turn out that
\bbb
T(w_1)X^\m(w_2)=\frac{\p X^\m(w_2)}{w_{12}}+({\rm smooth \ terms})+{O}(|X|^{-3})\ .
\eee
To prove this, we first investigate the OPE of $T^{[\beta^0]}$ with $X$, where we will see that
\bbb
\left.T^{[\beta^0]}(w_1)X^\m(w_2)\right|_{\rm prop.\ correction} = 
- T^{[\beta^1]}(w_1)X^\m(w_2) + ({\rm smooth\ terms})+{ O}(|X|^{-3})\ .
\eee
The contraction of $T^{[\beta^0]}$ with the PS vertex $-S^{[\beta^1]}=-\int d^2w {\mathcal L}_{\rm PS}$ \SilentOthersComment{(in the conformal gauge. the negative sign comes from the exponent of $e^{-S}$)} is
\be
T^{[0]}(w_1)\cdot (-S^{[\beta]})
&=& 2\beta\p X_\mu \p X^\mu\frac{\p\bp\phc}{\inv_{11}}(w_1)
\nn\\
&&\kern-30pt -\frac\beta{\pi}\int d^2w_3 
\frac{\p X_\mu(w_1)\bp X^\mu(w_3)}{w_{13}^2}\frac{\p\bp\phc}{\inv_{11}}(w_3)+{\mathcal O}(|X|^{-2})\ .
\ee
By partial integration, and using the leading order EOM, the second term in the RHS becomes
\begin{align*}
\int d^2w_3 
&\frac{\p X_\mu(w_1)\bp X^\mu(w_3)}{w_{13}^2}\frac{\p\bp\phc}{\inv_{11}}(w_3)\\
&=\int d^2w_3 \left[\p^2\phc(w_3)+(\p\phc(w_3))^2\right]
\bp_2\left[\frac{1}{w_{13}}\frac{\p X_\mu(w_1)\bp X^\mu(w_3)}{\inv_{11}(w_3)}\right]
\\
&=-2\pi\left[\p^2\phc(w_1)+(\p\phc(w_1))^2\right]
\\&+\int d^2w_3 \left[\p^2\phc(w_3)+(\p\phc(w_3))^2\right]\sum_{n=1}^\infty\frac{w_{13}^{n-1}}{n!}\bp\frac{\p^n\inv_{11}}{\inv_{11}}(w_3)\ .
\end{align*}
The latter term can be neglected because when contracting with $X^\mu(w_2)$, it does not yield 
any singularity in the limit $w_{1}-w_2\to 0$ . Finally,
\be
T^{[0]}(w_1)\cdot (-S^{[\beta]}) 
&=& T^{[\beta]}(w_1)
\nn\\
&& +{(\rm terms \ which \ are \ negligible \ in \ the \ above\ sense)}
\nn\\
&& +{\mathcal O}(X^{-2})\ .
\ee
Contracting with $X^\mu(w_2)$, we get
\bbb
\left.T^{[0]}(w_1)X^\mu(w_2)\right|_{\rm prop.\ correction}=
-\left.T^{[\beta]}(w_1)X^\mu(w_2)\right.+ ({\rm smooth})+{ O}(|X|^{-3})\ .
\eee
The result is that $T\ups{\b\uu 0 + \b\uu 1}(w\ll 1) \cc X\uu\m(w\ll 2) =   {{1}\over {w\ll{12}}} \cc \pp X\uu\m(w\ll 2)
+ ({\rm smooth}) + O(|X|\uu{-3})$, where the contractions are performed with the full propagator.  
To put it more concisely,
\bbb
T\ups{{\rm first~two~orders~in~}\b}(w\ll 1)\cc X\uu\m(w\ll 2) =   {{1}\over {w\ll{12}}} \cc \pp X\uu\m(w\ll 2)
+ ({\rm smooth}) + O(|X|\uu{-3})\ ,
\een{TXOPENonRenormEq}
where the OPE is performed in the full quantum theory.  
In other words, the OPE of $T\ups{{\rm full}}$ with $X$ in the full interacting theory
is the same as that of the free $T\ups{\b\uu 0}$ in the free theory, modulo terms of order $|X|\uu{-3}$ and smaller.

This proves the standard OPE of $T$ with $X$ is unmodified by interaction and quantum corrections up to and including
relative order $\b\cc |X|\uu{-2}$.  In other words, to first subleading order, the conformal transformation of $X$ 
remains that of a primary field of weight zero.   This was inevitable, as we formulated our 
theory from the outset with explicit Weyl invariance built in, from which
the conformal invariance of the gauge-fixed theory is inherited.  The operators in the CFT 
describing the theory in conformal gauge receive
their conformal properties from the Weyl transformations of the fields in the path integral.  
Since our construction of the Weyl-invariant quantum theory
was based on assigning $X$ to be invariant under Weyl transformations, it automatically appears as a 
conformal primary of weight zero in the interacting CFT.

\subsection{Closure of the OPE of $T$ with itself}

We now demonstrate that the OPE of $T$ with itself satisfies the standard form, with $c=D+12\beta=26$, 
up to and including ${\mathcal O}(|X|^{-1})$,
using the result of the previous section. First, when we separate $X^\mu=E^\mu+Y^\mu$, 
where $E^\mu$ is a $c$-number solution of $\p\bp E^\mu=0$ and $Y^\mu$ is an operator 
representing fluctuations, the $TX$ OPE is the same as the $TY$ OPE:
\begin{equation}\label{TYOPE}
\left.T^{[\beta^0+\beta^1]}(w_1)Y^\mu(w_2)\right.\sim\frac{1}{w_{12}}\p (E^\mu+Y^\mu)(w_2)+{\mathcal O}(E^{-3})\ .
\end{equation}
Then, expanding $T$ with respect to $Y$,
\be
T^{[\beta^0+\beta^1]}&=& \left.T^{[\beta^0+\beta^1]}\right|_{X=E}-\frac1\al\left(2\p E\cdot\p Y + \p Y\cdot \p Y\right) 
\nn\\
&&\kern-50pt +\beta\left[-\p E\cdot\p E\left(\frac{\p E\cdot\bp Y+\p Y\cdot \bp E}{\eii^2}\p\bp\log\eii-\frac1\eii\p\bp\frac{\p E\cdot\bp Y+\p Y\cdot \bp E}{\eii}\right)\right.
\nn\\
&&
\kern-50pt \left.+2\p E\cdot\p Y\frac{\p\bp{\rm ln}\eii}{\eii}+\p^2\frac{\p E\cdot\bp Y+\p Y\cdot \bp E}{\eii^2}-\p{\rm ln}\eii\p\frac{\p E\cdot\bp Y+\p Y \cdot\bp E}{\eii^2}\right]
\nn\\
&&
\kern-50pt +{\mathcal O}(Y^2E^{-2})\ ,
\ee
with $E\ll{11}$ as in \rr{EpqDef}.
A tedious but straightforward calculation using eqn.~(\ref{TYOPE}) gives
\be
T^{[\beta^0+\beta^1]}(w_1)T^{[\beta^0+\beta^1]}(w_2) &\sim & \frac{D+12\beta}{2w_{12}^4}+\frac{2}{w_{12}^2}T^{[\beta^0+\beta^1]}(w_2)
\nn\\
&&+\frac1{w_{12}}\p T^{[\beta^0+\beta^1]}(w_{2}) +{\mathcal O}(|X|^{-2})\ .
\ee
This OPE implies that the theory is a conformal field theory with $c=26$, regardless of the space-time dimensions and the form of the classical solution.  This result generalizes the stress tensor OPE in \cite{Polchinski:1991ax}
in the special case where $E\uu\m$ is linear in the worldsheet coordinates.


\subsection{OPE of $T$ with composite operators}\label{TCompositeOPE}
Suppose ${\cal O}$ is some operator with definite or polynomial $|X|$-scaling $|X|\uu p$, and we would like to compute the OPE of
$T$ with ${\cal O}$ up to (and excluding) terms of order $|X|\uu{-4}$, relative to ${\cal O}$ 
itself.\footnote{That is, we would like to compute
the OPE of $T$ with ${\cal O}$ up to (and excluding) terms of order $|X|\uu{p-4}$.} 

Terms involving cubic or higher interaction vertices are too small to contribute. 
To connect $T$ and ${\cal O}$ with a cubic
interaction vertex, we would need to pull three $X$ fields from $T$ and ${\cal O}$, and the cubic interaction vertex from
the PS term itself scales as $|X|\uu{-3}$.  Thus, the total effect scales as $|X|\uu{-6}$, 
relative to the classical combined $X$-scaling
of $T{\cal O}$, the latter being $|X|\uu{p+2}$.  So any contribution to the $T{\cal O}$ OPE involving a cubic vertex is no larger
than $|X|\uu{p-4}$.  Likewise, a diagram with a $q\uth$-order vertex, with $q \geq 4$, must connect the two operators by pulling off
at least one $X$ field from each, and the vertex itself scales as $|X|\uu{-q}$, so the $|X|$-scaling of the contribution is
$|X|\uu{-q-2}$, relative to the classical $|X|$-scaling $|X|\uu{p+2}$ of the classical operator product $T{\cal O}$.  
The $|X|$-scaling of the term is thus $|X|\uu{p-q}$ or smaller, and thus no contribution with a quartic or higher interaction vertex will
contribute to the OPE at the desired order of accuracy.

Therefore, the only terms contributing to the $T{\cal O}$ OPE up to the order of interest are classical propagators, with or without
interaction corrections.  For a single propagator, we have found that the effect of the propagator correction on the singular
terms in the OPE precisely cancels
the effect of the explicit order-$\b\uu 1 |X|\uu 0$ correction to the stress tensor.  
For two propagators, a double contraction with
leading-order propagators is order-$|X|\uu{-4}$, relative to the classical operator product, 
and each interaction correction to a propagator
suppresses the $|X|$-scaling further by $|X|\uu{-2}$.  
A single interaction correction to either propagator is already order-$|X|\uu{p-4}$ in
total, and thus small enough to ignore.  Likewise, contractions using three or more propagators, 
even uncorrected propagators, is already order-$|X|\uu{p-4}$, and can be ignored as well.

Thus, the OPE of $T\ups{{\rm full}}$ with ${\cal O}$ is given by a sum of the following:
\bi
\item{Terms coming from contractions with a single (corrected) propagator, which implement a classical conformal transformation with $X$
transforming as a scalar that is invariant under Weyl transformations;}
\item{An anomalous term coming from a double contraction using free propagators;}
\item{Terms of $X$-scaling $|X|\uu{p-4}$ or smaller;}
\item{Smooth terms.}
\ei
For purposes of classifying conformal primary operators at next-to-leading order in the effective string expansion, we care about
the first two contributions only.

\section{Conclusions}

We have embedded effective string theory in the Polyakov formalism, 
so that the standard worldsheet diffeomorphism and Weyl symmetries are manifest.  
The absence of the Weyl anomaly is clearly proven prior to gauge fixing.  
The classification of diffeomorphism-invariant operators with definite Weyl scaling implies that no 
adjustable parameter exists at next-to-leading order, whereas
at next-to-next-to-leading order the action is not universal.

One of the efficiencies of our formalism, relative to the old covariant one, is that the primary 
conformal transformations of $X$ are maintained in the presence of the interaction, by construction.  
Thus, to classify conformal primaries at next-to-leading order, only a classical conformal transformation 
and a double contraction using free propagators need be taken into account. 

\section*{Acknowledgments}
\addcontentsline{toc}{section}{Acknowledgments}
SH is grateful to the Weizmann institute for hospitality while this work was in progress, and 
would also like to thank J. Sonnenschein, O. Aharony, and M. Field for extremely useful discussions during
his visit.   JM thanks the Stanford Institute for Theoretical Physics for hospitality during the completion of this paper. 
The work of SH, SM, and JM is supported by
the World Premier International Research Center Initiative
(WPI Initiative), MEXT, Japan.  The work of SH was also supported in part
by Grant-in-Aid for Scientific Research (22740153) and Grant-in-Aid for Scientific Research (26400242) from the Japan Society for Promotion of Science (JSPS).


\begin{thebibliography}{99}


\bibitem{Polchinski:1991ax} 
  J.~Polchinski and A.~Strominger,
  ``Effective string theory,''
  Phys.\ Rev.\ Lett.\  {\bf 67}, 1681 (1991).


\bibitem{Aharony:1999ti} 
  O.~Aharony, S.~S.~Gubser, J.~M.~Maldacena, H.~Ooguri and Y.~Oz,
  ``Large N field theories, string theory and gravity,''
  Phys.\ Rept.\  {\bf 323}, 183 (2000)
  [hep-th/9905111].


\bibitem{Aharony:2013ipa} 
  O.~Aharony and Z.~Komargodski,
  ``The Effective Theory of Long Strings,''
  JHEP {\bf 1305}, 118 (2013)
  [arXiv:1302.6257 [hep-th]].

\bibitem{natsuume} 
  M.~Natsuume,
  ``Nonlinear sigma model for string solitons,''
  Phys.\ Rev.\ D {\bf 48}, 835 (1993)
  [hep-th/9206062].





\bibitem{Luscher:1980fr} 
  M.~Luscher, K.~Symanzik and P.~Weisz,
  ``Anomalies of the Free Loop Wave Equation in the WKB Approximation,''
  Nucl.\ Phys.\ B {\bf 173}, 365 (1980).

\bibitem{Luscher:1980ac} 
  M.~Luscher,
  ``Symmetry Breaking Aspects of the Roughening Transition in Gauge Theories,''
  Nucl.\ Phys.\ B {\bf 180}, 317 (1981).
  

\bibitem{Luscher:2004ib} 
  M.~Luscher and P.~Weisz,
  ``String excitation energies in SU(N) gauge theories beyond the free-string approximation,''
  JHEP {\bf 0407}, 014 (2004)
  [hep-th/0406205].
  

\bibitem{HariDass:2007gn} 
  N.~D.~Hari Dass and P.~Matlock,
  ``Covariant Calculus for Effective String Theories,''
  arXiv:0709.1765 [hep-th].
  
  
\bibitem{Dubovsky:2012sh}   
  S.~Dubovsky, R.~Flauger and V.~Gorbenko,
  ``Effective String Theory Revisited,''
  JHEP {\bf 1209}, 044 (2012)
  [arXiv:1203.1054 [hep-th]].
  

\bibitem{Luscher:2002qv} 
  M.~Luscher and P.~Weisz,
  ``Quark confinement and the bosonic string,''
  JHEP {\bf 0207}, 049 (2002)
  [hep-lat/0207003].



\bibitem{Aharony:2010db} 
  O.~Aharony and N.~Klinghoffer,
  ``Corrections to Nambu-Goto energy levels from the effective string action,''
  JHEP {\bf 1012}, 058 (2010)
  [arXiv:1008.2648 [hep-th]].


\bibitem{Aharony:2011ga} 
  O.~Aharony, M.~Field and N.~Klinghoffer,
  ``The effective string spectrum in the orthogonal gauge,''
  JHEP {\bf 1204}, 048 (2012)
  [arXiv:1111.5757 [hep-th]].





\bibitem{Aharony:2009gg} 
  O.~Aharony and E.~Karzbrun,
  ``On the effective action of confining strings,''
  JHEP {\bf 0906}, 012 (2009)
  [arXiv:0903.1927 [hep-th]].
\bibitem{Aharony:2010cx} 
  O.~Aharony and M.~Field,
  ``On the effective theory of long open strings,''
  JHEP {\bf 1101}, 065 (2011)
  [arXiv:1008.2636 [hep-th]].
\bibitem{Aharony:2011gb} 
  O.~Aharony and M.~Dodelson,
  ``Effective String Theory and Nonlinear Lorentz Invariance,''
  JHEP {\bf 1202}, 008 (2012)
  [arXiv:1111.5758 [hep-th]].
\bibitem{AharonyUnPub}
  O.~Aharony, Z.~Komargodski and A.~Schwimmer, presented by O.~Aharony
  at the Strings 2009 conference, June 2009,
  http://strings2009.roma2.infn.it/talks/Aharony\_Strings09.ppt .



\bibitem{Baker:2000ci} 
  M.~Baker and R.~Steinke,
  ``Effective string theory of vortices and Regge trajectories,''
  Phys.\ Rev.\ D {\bf 63}, 094013 (2001)
  [hep-ph/0006069].

  
\bibitem{Baker:2002km} 
  M.~Baker and R.~Steinke,
  ``Semiclassical quantization of effective string theory and Regge trajectories,''
  Phys.\ Rev.\ D {\bf 65}, 094042 (2002)
  [hep-th/0201169].



\bibitem{Kruczenski:2004me} 
  M.~Kruczenski, L.~A.~Pando Zayas, J.~Sonnenschein and D.~Vaman,
  ``Regge trajectories for mesons in the holographic dual of large-N(c) QCD,''
  JHEP {\bf 0506}, 046 (2005)
  [hep-th/0410035].



\bibitem{Sonnenschein:2014jwa} 
  J.~Sonnenschein and D.~Weissman,
  ``Rotating strings confronting PDG mesons,''
  arXiv:1402.5603 [hep-ph].
  
\bibitem{Hellerman:2013kba} 
  S.~Hellerman and I.~Swanson,
  ``String Theory of the Regge Intercept,''
  arXiv:1312.0999 [hep-th].




\bibitem{Polchinski:1998rq} 
  J.~Polchinski,
  ``String theory. Vol. 1: An introduction to the bosonic string,''
  Cambridge, UK: Univ. Pr. (1998) 402 p



















































  


  
  
  
  



  
  

    
  
  
  

  
  
  
  
  

  
  
  
    
  
  

  
  

  
  
  
  
  
  
  













 




\bibitem{Drummond:2004yp} 
  J.~M.~Drummond,
  ``Universal subleading spectrum of effective string theory,''
  hep-th/0411017.

\bibitem{Dubovsky:2013gi} 
  S.~Dubovsky, R.~Flauger and V.~Gorbenko,
  ``Evidence for a new particle on the worldsheet of the QCD flux tube,''
  arXiv:1301.2325 [hep-th].









  



\bibitem{Wilczek:2004im} 
  F.~Wilczek,
  ``Diquarks as inspiration and as objects,''
  In *Shifman, M. (ed.) et al.: From fields to strings, vol. 1* 77-93
  [hep-ph/0409168].




  

  
\end{thebibliography}
\end{document}